\documentclass[aps,pra,amsmath,amssymb,twocolumn,showpacs,nofootinbib]{revtex4}

\usepackage{graphicx}
\usepackage{dcolumn}
\usepackage{bm}
\usepackage{lipsum}

\usepackage{epsfig}
\usepackage{colordvi}
\usepackage{float}
\usepackage{accents}

\newcommand{\be}{\begin{equation}}
\newcommand{\ee}{\end{equation}}
\newcommand{\ba}{\begin{eqnarray}}
\newcommand{\ea}{\end{eqnarray}}
\newcommand{\n}{\nonumber}

\begin{document}
\title{Direct Measurement of Time-Frequency Analogues of Sub-Planck Structures}

\author{Ludmila Praxmeyer$^1$, Chih-Cheng Chen$^1$, Popo Yang$^2$, Shang-Da Yang$^1$,  and Ray-Kuang Lee$^{1,2}$}

\affiliation{
$^1$ Institute of Photonics Technologies, National Tsing-Hua University, Hsinchu, Taiwan\\
$^2$ Department of Physics, National Tsing-Hua University, Hsinchu, Taiwan}


\begin{abstract}
Exploiting the correspondence between Wigner distribution function
and a frequency-resolved optical gating (FROG) measurement, we
 experimentally demonstrate existence of the chessboard-like
interference patterns  with a time-bandwidth product smaller than
that of a transform-limited pulse in the phase space  representation
of compass states. Using superpositions of four electric pulses as
realization of compass states, we have shown via direct measurements
that displacements leading to orthogonal states can be smaller than
limits set by uncertainty relations. In the experiment we observe an
exactly chronocyclic correspondence to the sub-Planck structure in
the  interference pattern appearing for superposition of two
Sch\"{o}dinger-cat-like states in a position-momentum phase space.
\end{abstract}

\maketitle

It is the superposition principle that leads to interference and
diffraction phenomena, determines evolution of wavepackets in
classical and quantum systems.
When applying to macroscopic objects, the interpretation paradox of wave function in superposition arises from  Schr\"{o}dinger's {\sl gedanken} experiment on cat states~\cite{Schrodinger-cat}.
By a Schr\"{o}dinger-cat-like state in quantum optics one usually
understands a coherent superposition of two coherent states, say,
{\small$|\alpha\rangle+|-\alpha\rangle$}. Experimentally, such
superpositions were created, e.g., in the atomic and molecular
systems~\cite{ion, Haroche},  superconducting
circuits~\cite{supQbits, Wal}, and quantum optical
setups~\cite{Ourjoumtsev-1, Tak}.
 It has been noted by Zurek~\cite{Zurek} that  a superposition of two  Schr\"{o}dinger-cat-like states
(four coherent states in total,
{\small$|\alpha\rangle+|-\alpha\rangle+|i\alpha\rangle+|-i\alpha\rangle$},
a so called {\it compass state}) in a Wigner phase-space description
\cite{wigner1932} gives rise to the interference structure changing
rapidly on an area smaller than a
  Planck's constant $\hbar$. The result is counterintuitive because the
  Heisenberg's uncertainty principle sets a limitation on
the simultaneous resolutions in two conjugate observables.

It turns out that the sub-Planck structure determines scales
important to the distinguishability of quantum states~\cite{Zurek},
thereby, potentially has an impact on an ultra-sensitive quantum
 metrology~\cite{ion-trap, Toscano} and could affect efficient storage of quantum
 information~\cite{entangle-2, Science}.
In principle, these sub -Planck structures could be used to improve the sensitivity in
weak-force detection~\cite{Roy, weak} and  to help maintaining a
high fidelity in the continuous-variable teleportation
protocols~\cite{decay, teleportation}.  In practice, the classical wave optics
 analogues of sub-Planck structures in time-frequency
 domain were observed experimentally so far  only for
superposition of two Gaussian pulses~\cite{Ludmila, Walmsley}.
Obviously, a single Schr\"{o}dinger-cat-like state offers a
sensitivity to the perturbations only in  one direction:
perpendicularly to the line joining the coherent states. To provide
sensitivity in all directions, another pair of  coherent states is
needed. Theoretical proposals for generation of compass states
include interaction in cavity-QED systems~\cite{GSA-PRA,Science},
evolution in  a Kerr medium~\cite{Tanas, Magda} and fractional
revivals of molecular wavepackets~\cite{Ghosh}.
Nevertheless, there was no  experimental proof in the more complex and demanding case
of superposition of four pulses.
The main technical challenge in the preparation of states separated
simultaneously in two conjugate coordinates comes from the
difficulty in keeping the coherence among them. Furthermore,
performing measurements of such tiny phase space structures is far
from trivial.

In this Letter, we report a direct measurement of a
compass state superposition of four optical pulses and the
corresponding interference pattern
in the time-frequency domain.
 The mathematical equivalence between the electric field of an ultrashort pulse and
the quantum mechanical wave function of the same shape, enables us to
observe the phase space structures through the time-dependent
spectrum of light. The experimental data obtained from
frequency-resolved optical gating (FROG) measurements of light
pulses reveals sub-Planck structure analogues corresponding directly
to the compass states. In the interference patterns, areas smaller
than that of transform-limited pulses are measured, illustrating
that displacements
 leading to orthogonality of compass states can be smaller  than the
 Fourier limit imposed on the pulses forming these  superpositions.

\begin{figure}[H]
\begin{flushleft}
(b)
 \vspace*{-.7cm}
 $\;$\includegraphics[width=7.6cm]{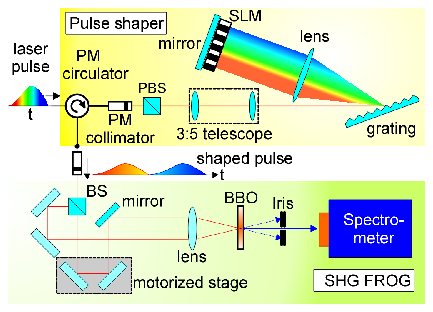}\\
 \vspace*{-2.6cm}(a)\vspace*{2.6cm}
\end{flushleft}
  \caption{(Color online) Schematic view of our experimental setup, including
  (a) Pulse shaper composited of a set of grating, lens, and spatial light
  modulator (SLM); (b) SHG FROG
  composited by a beam splitter (BS) and a motorized stage in one arm to
  control time delay $\tau$,  a  nonlinear crystal (BBO)
   for the second harmonic generation (SHG), and a
   spectrometer.}\label{f1}
\begin{center}
\includegraphics[width=8cm]{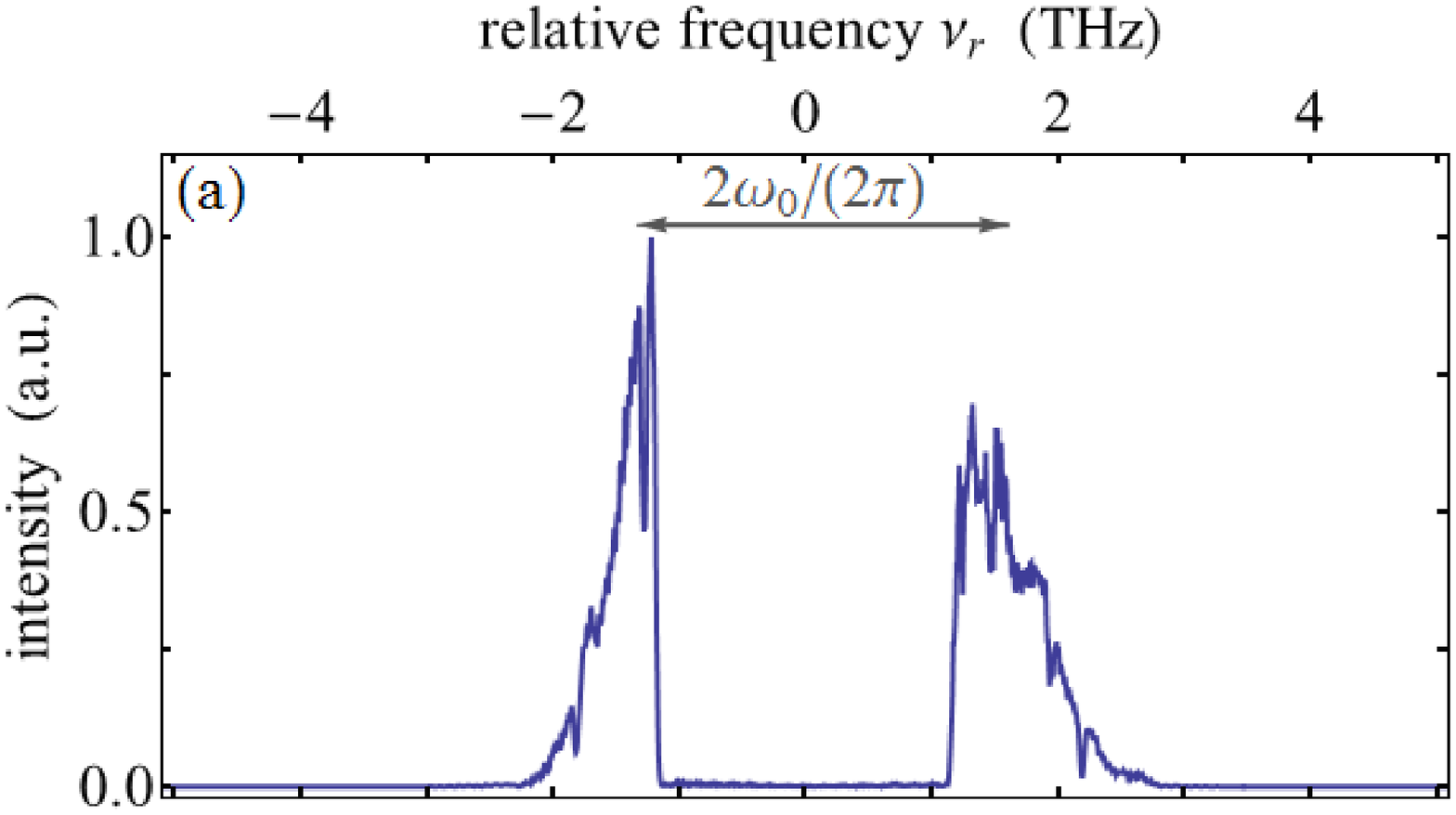}\\
{}\vspace*{-.8cm}\includegraphics[width=8cm]{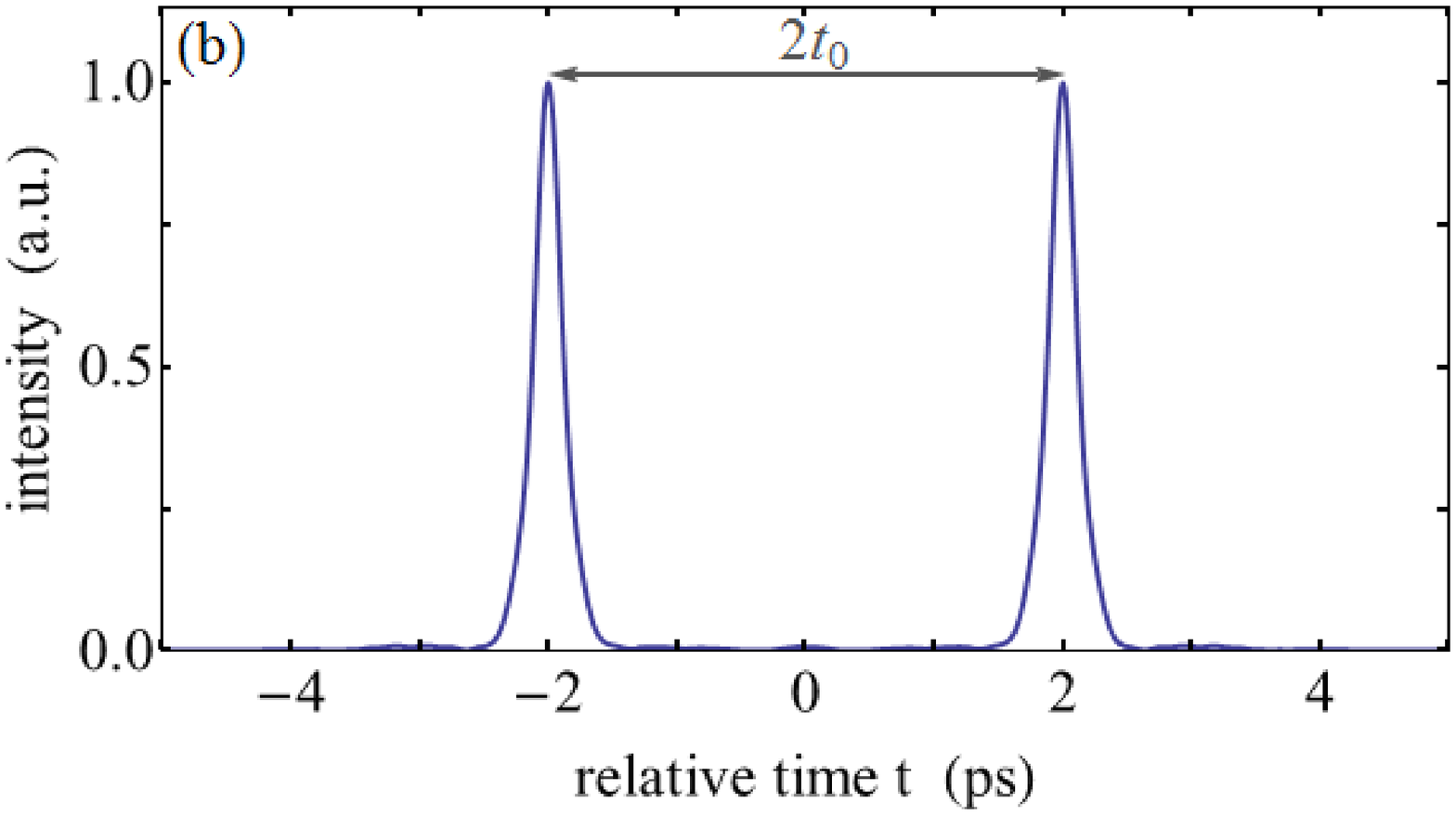}
\end{center}
{}\vspace*{-.4cm}
\caption{Typical experimental profiles for our compass states
are shown as a function of (a) relative frequency
$\nu_r=\omega/(2\pi)$ and (b) time $t$.
 Here, the four pulses
 are  concurrently separated by $\omega_0/\pi=3.3\,$THz in frequency
and $2t_0=4\,$ps in time.}\label{f2}
\end{figure}

As illustrated in
Fig.~\ref{f1}(b), in our second-harmonic generation (SHG) FROG
measurement~\cite{frog-book} the input optical pulse $E(t)$ is split
into
 two replicas with a tunable time delay by passing through a
 standard Michelson interferometer.
These two mutually delayed replicas mix in a $\chi^{(2)}$ nonlinear
crystal such as the Barium borate (BBO) crystal used in our
experiment. Then, a spectrally resolved sum frequency signal is
recorded for each time delay $\tau$. The resulting time-frequency
map (spectrogram) implemented by the nonlinear self-optical gating
mechanism can be formulated as the following function for an initial
field $E(t)$:
\begin{eqnarray}
\label{frogSHG}
\mathcal{I}^{\text{FROG}}_{\text{E}}(\tau,\omega)
=\biggl|\int_{\scriptscriptstyle{-\infty}}^{\scriptscriptstyle{\infty}}\!\!
\!\! E(t)\, E(t-\tau)\,e^{i\omega t} \,dt\biggl|^2.
\end{eqnarray}
For pulses with real envelopes and a linear phase, the
spectrogram~(\ref{frogSHG}) is easily mapped to the
 Wigner  quasi-probability distribution $W(q,p)$~\cite{wigner1932}:
\begin{eqnarray}
\label{wigfunc}
W(q,p) =  \frac{1}{\pi \hbar}  \int_{-\infty}^{\infty}\,
e^{2 i\xi p/\hbar}\,F( q-\xi){F}^\ast( q+\xi) \,d\xi,
\end{eqnarray}
as
\begin{eqnarray}
\mathcal{I}^{\text{FROG}}_{\text{E}}(\tau, \omega) \propto
 \big| W(\tau/2, \omega/2) \big|^2.
 \end{eqnarray}
Here, ${F}^\ast(q)$ denotes the complex conjugate of $F(q)$, $\hbar$
is set to $1$, and our time and frequency domains are represented by $(\tau,\omega)$.
Moreover,  under these conditions, all cross-sections of the FROG
spectrogram correspond to a scalar product between the probe state
$E(t)$ and its `twin' appropriately shifted in time or
frequency~\cite{chirp, Ludmila}. Thus, zeros of the cross-sections
appear for these values of time and frequency shifts that lead to
orthogonal states. In the following, we check the scale of the phase
space displacements resulting in distinguishable states based on
this property. Let us stress that the data presented here comes from
direct spectrogram measurements and, unlike typical FROG
applications, does not involve steps of state reconstruction.

In our experiment, an  Er-doped fiber laser with $1564\,$nm
  center wavelength, $37\,$nm spectral bandwidth and $5.68\,\mathrm{MHz}$
  repetition rate is used as the light source.
To construct a compass state, i.e., the superposition of four coherent states,
 a pulse shaper is introduced to split the input pulse in both time
and frequency domains, as illustrated in Fig.\ref{f1}(a)~\cite{C-C}.
Here, a telescope is employed to improve the spectral resolution of
the pulse shaper by expanding   the beam diameter from
$2.8\,\mathrm{mm}$ (at the collimator) to $4.5\,$mm. The input pulse
is directly shaped in frequency domain, though a set of
 grating, lens, and a spatial light modulator (SLM).
The whole spectrum occupies $320$ pixels in the SLM, and the
throughput of our pulse shaper is around $30$\%. The spectral separation
is realized by blocking the central range of the input spectrum, see
Fig.~2(a); while the temporal separation $2t_0$ is achieved by
imposing an extra mask function
 $M_{\text{SLM}}(\omega) = \cos(\omega t_0)e^{- i \omega t_0}$
via the SLM~\cite{opt}. The spectrally and temporally separated
coherent pulses are further made transform limited by compensating
the residual spectral phase via the phase modulation function of the
pulse shaper. The power spectrum and temporal intensity of an
example are shown in Figs.~\ref{f2}(a) and ~\ref{f2}(b), respectively.  In total,
an initial laser spectra was divided into four pulses separated
simultaneously by $2\omega_0$ in frequency and $2t_0$ in time, i.e.,
shaped into our time-frequency representation of the compass state.
For practical reasons, in the experiment  separation between pulses
in frequency was  kept constant while time separation $2t_0$ varied
from $1.5\,$ps to $5\,$ps.

In the ideal case, when the pulses cut from the initial laser
spectrum have Gaussian envelopes, the state generated in a pulse
shaper has the form
\begin{eqnarray}
E_{\text{in}}(t) &\simeq & e^{-(t-t_0)^2-i(\omega -\omega_0)
t}+e^{-(t+t_0)^2-i(\omega-\omega_0)
t}\n\\
&+&e^{-(t-t_0)^2 -i(\omega+\omega_0)t }   + e^{-(t+t_0)^2
-i(\omega+\omega_0)t },\label{kwadrat}
\end{eqnarray}
\noindent
where $2t_0$ and $2\omega_0$ are the time and frequency separations
between the Gaussian peaks. For the clarity of notation,
normalization coefficients and parameters denoting width of the
 wavepackets are omitted in Eq.~(\ref{kwadrat}). Depending on the context, in the text that follows
we will use either a regular frequency $\nu$ or the corresponding
angular frequency $\omega=2\pi\nu$.  The compass state
$E_{\text{in}}(t)$ from Eq.~(\ref{kwadrat})
 is constructed as a superposition of two mutually delayed pairs of pulses with the same
carrier frequency in each pair; it  corresponds to the original
compass state from Zurek paper~\cite{Zurek} rotated by $\pi/{4}$ in
the phase space.

\begin{figure}[t]
\begin{center}
\includegraphics[width= 8.4cm]{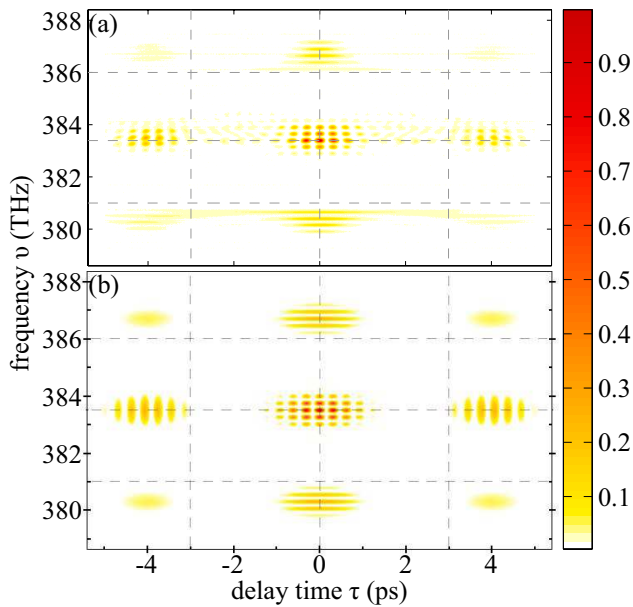}
\end{center}
\caption{(Color online) Compass state in the phase space obtained by (a) FROG map
measured for  $t_0=2$ps; and (b) numerical simulation calculated for
a superposition of four perfect identical Gaussian
pulses.}\label{4ps}
\end{figure}
A comparison between a theoretical simulation  and experimentally
obtained time and frequency phase space maps of  compass states is
presented in Fig.~\ref{4ps}, where spectrograms corresponding to the
same time and frequency separation between the cat states are
plotted. Experimental data from SHG FROG spectrogram measured for
$t_0=2\,$ps is shown in Fig.~\ref{4ps}(a); while the theoretical
plot calculated analytically for perfect Gaussian  pulses is
presented in Fig.~\ref{4ps}(b). In addition to four peaks
representing Gaussian wavepackets and located at the four corners of
the plots,  characteristic patterns of interference fringes are
clearly visible between every two of the peaks.  Moreover, in the
center of four perimeters,  a chessboard-like interference
pattern appears. In the following, we will verify that for large
enough separation between the peaks this chessboard-like pattern is
constructed from areas smaller than $\Delta
\tau \Delta \omega=1/2$ indicated by uncertainty relation, which
would correspond to sub-Planck areas in the position-momentum phase
space. For the shape of our pulses is obtained by cutting the laser
spectra, a slight discrepancy between the patterns formed by
the real experimental profiles and theoretically  considered
ideal Gaussian pulses can be seen between  Figs.~\ref{4ps}(a)
and~\ref{4ps}(b).

\begin{figure}[t]
\begin{center}
\includegraphics[width=8.4cm]{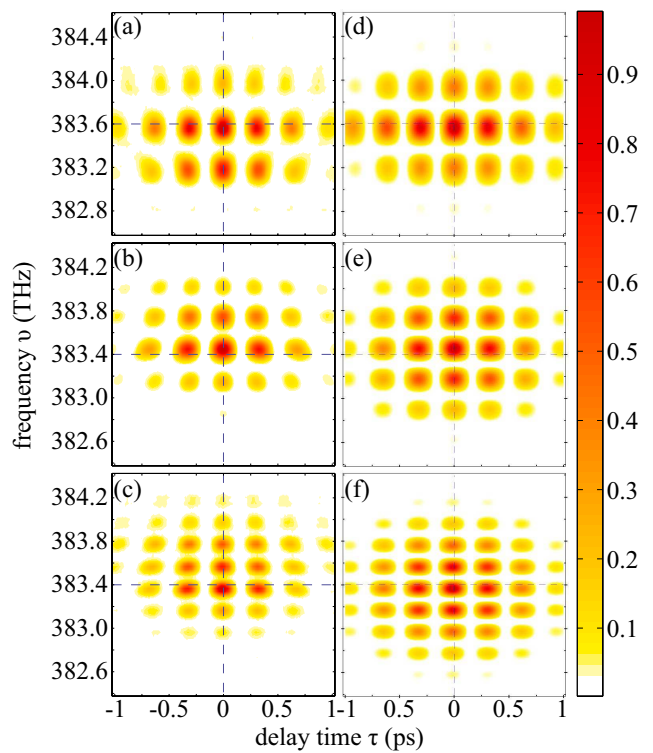}
\end{center}
\caption{(Color online) Zooms of the central interference structure of
FROG maps measured for (a) $t_0=1.25$ps, (b) $t_0=1.75$ps,  and (c)
$t_0=2.5$ps. As a comparison, the corresponding simulations
calculated for compass states build from perfect Gaussian pulses are
shown in (d), (e), and (f), respectively.}\label{f4}
\end{figure}
In the Wigner representation of a compass state, the middle
interference pattern  is build up from small rectangles of
alternating positive and negative values of the function. Even
though recordings in the FROG spectrograms take on only
non-negative values,
 areas of the above-mentioned rectangles remain the same, i.e.,  the distances between  subsequent
 zero lines do not change.
 For a large enough separation distances between pulses forming the superposition,
 these areas
 grow smaller than the area unit defined by the elementary uncertainty relation.
In quantum mechanical version, the area unit is equal to
$\hbar/2$; while in a chronocyclic phase space, it is simply
$1/2$.  It is worth noting that even for smaller separation
distances ``sub-Fourier" areas might appear in the FROG maps,
namely, the areas  smaller than those corresponding to dispersion
 $\Delta \tau \Delta \omega$ calculated  for any from the pulses  forming the
compass state superposition. To demonstrate  how  the interference
structure appearing in the middle of FROG maps changes with change
of initial parameters, in Fig.~\ref{f4}, zooms of the central
parts of spectrograms for different time separations between pair
of pulses are presented. Figure~\ref{f4} shows the Zoom of the
central interference patterns obtained from the FROG traces for
(a) $t_0=1.25\,$ps, (b) $t_0=1.75\,$ps,  and {\mbox{(c)
$t_0=2.5\,$ps.}} The corresponding theoretical plots made for
superpositions of four identical Gaussian pulses  are shown in in
Figs.~\ref{f4}(d), (e), and (f), respectively. The comparison
serves to illustrate that areas between zero lines of  FROG maps
indeed decrease with an increasing separation between pair of
pulses. Again, as due to imperfections, amplitudes of pulses used
in the  experiment were not the same and the resulting
interference  patterns measured are not symmetric in respect to
the central wavelength line.

\begin{figure}[t]
\begin{flushleft}
 \includegraphics[width=8cm,height=5.5cm]{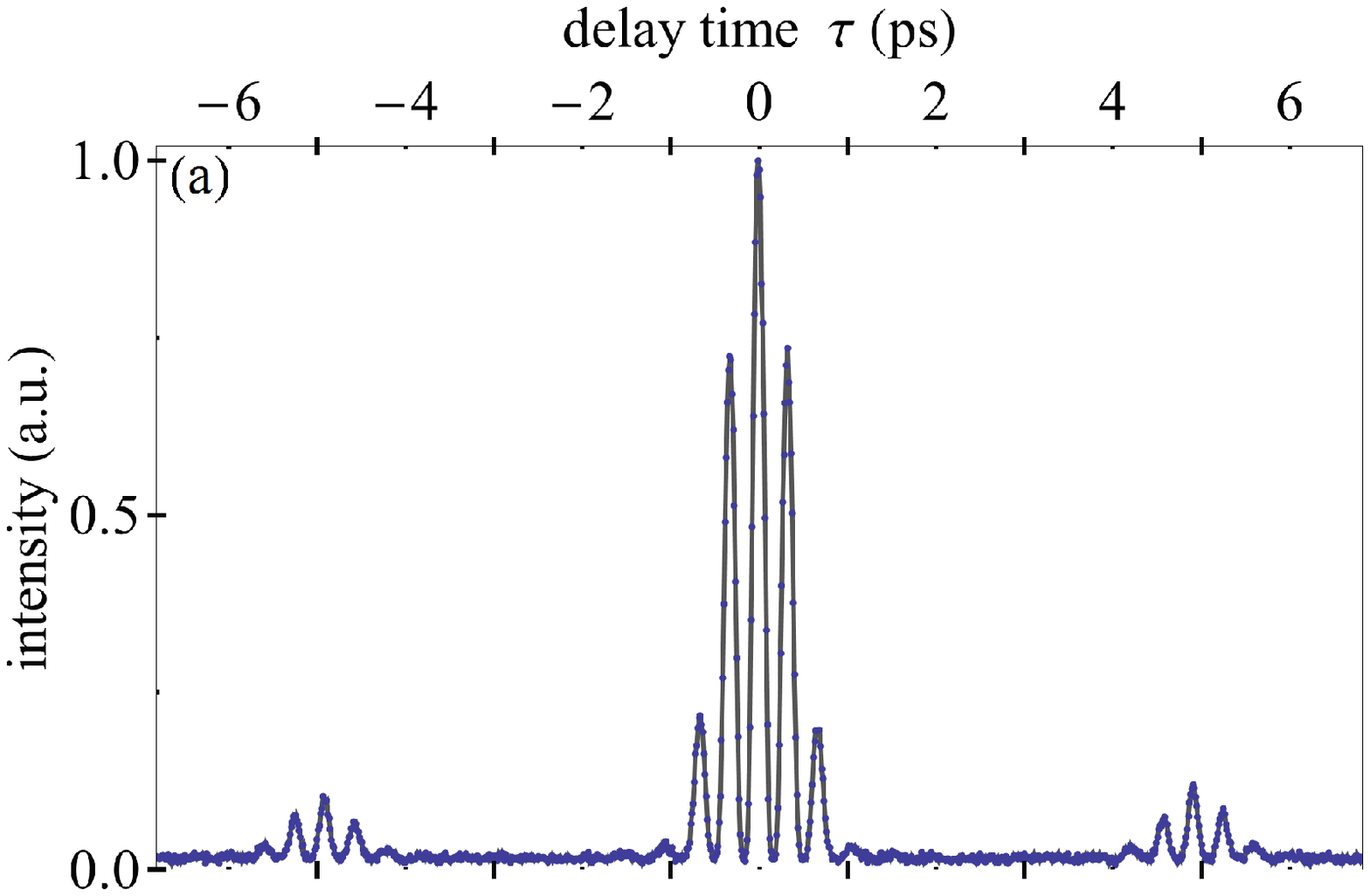} \\
\vspace*{-.75cm}\includegraphics[width=8cm,height=6cm]{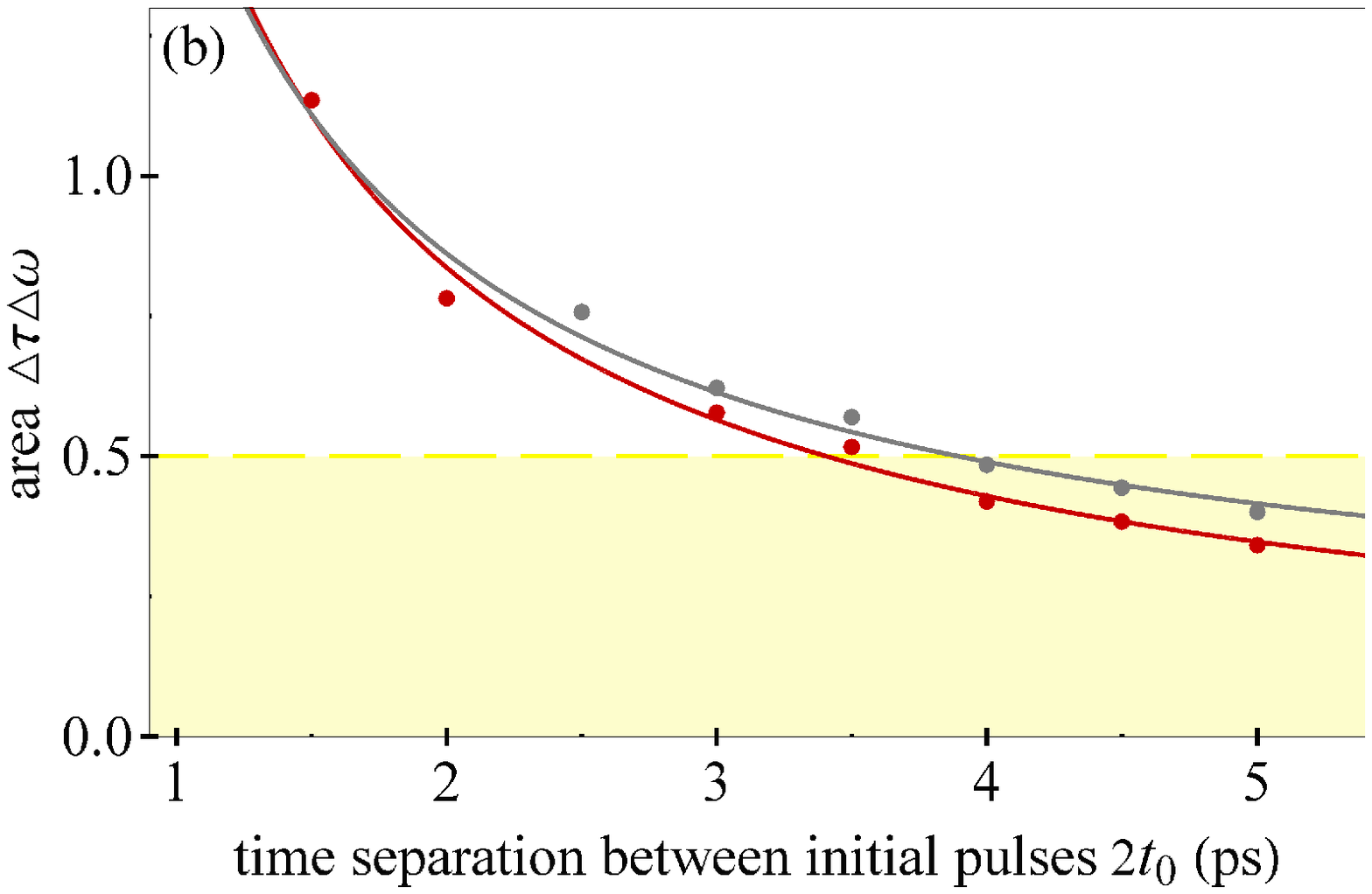}
\end{flushleft}
\vspace*{-.75cm}
 \caption{(Color online) (a)  A central frequency ($\nu= 383.36\,$THz)
 cross-section of the FROG map measured for $t_0=2.5\,\mathrm{ps}$. Here, blue dots depict
 measured intensity values for a given time delay $\tau$ introduced
 between two input state copies in the arms of FROG
 apparatus.
 (b) Average areas, $\Delta \tau \Delta \omega$, of the
 rectangles appearing between the zero lines in the interference structure of the central part of
 FROG maps.  Values depicted by red or gray dots correspond to two independent sets of experimental data, respectively.
 Region below the uncertainty relation limit of $0.5$ is denoted in yellow.}
 \label{fit}
\end{figure}

Finally, let us examine the cross-section of the FROG map measured
for $t_0=2.5\,$ps along the central wavelength $\lambda=782\,$nm
($\nu= 383.36\,$THz) presented in Fig.~\ref{fit}(a). As mentioned
before, the  cross-sections of FROG spectrograms give values of
the scalar product of a probe field with its ideal copy, but
shifted in phase space in a direction perpendicular to the
direction defining the cross-section. It is clearly seen that the
values of the function plotted in Fig.~\ref{fit}(a) decrease to
zero between the subsequent peaks. A time shift equal to one half
of the distance between the zeros results in the superpositions
orthogonal to the initial compass state. This result is
complementary to the one reported in Ref.~\cite{Ludmila}, where
zeros in  cross-sections defined by  $\tau=0$ were shown.
Figure~\ref{fit}(b) demonstrates how average areas  $\Delta \tau
\Delta
\omega$
 of individual rectangles forming the central interference pattern,  change with the
 change of parameter $t_0$.
The average values of $\Delta \tau \Delta \omega$ plotted in
Fig.~\ref{fit}(b) are calculated  for two independently collected
 sets of data (depicted by gray or red dots, respectively). These two sets of data are measured
for slightly different pulse shapes  and differently set time delay
resolution, i.e., different values of the smallest stage-motor step.
It is clearly seen that in both cases  for
larger separation distances, areas between zeros indeed reach below
limit imposed by uncertainty relation.
Exactly these areas determine a scale of smallest change ensuring
distinguishability of the mutually shifted states. As a last point,
we would like to underline that the uncertainty relation is  not
violated. What we prove here is that a sub-Fourier change of initial
state (shift in frequency and/or time) is enough to produce state
orthogonal to the initial superposition. In the quantum mechanics
this  leads to a perfect distinguishability
of states.

In summary, using  correspondence between FROG maps in a
chronocyclic phase space and the Wigner distribution function, we
have demonstrated existence of the interference structure changing
on areas smaller than that of a transform-limited pulse. We have
performed FROG measurements of compass states realized through
superpositions of four light pulses constructed with a help of a
spatial light modulator (SLM), by manipulating the phase difference in
the spectrum. Interpretation of the FROG spectrograms as  maps of
the values of a scalar product between probe pulses and  their
shifted copies, allowed us to show explicitly that the scale of
changes leading to orthogonality of states indeed goes below  the
Fourier limit if sufficient time and frequency separations are
simultaneously kept between the input pulses.  The sub-Planck
structure of the  interference pattern in phase space representation
of compass states not only manifests itself as a generalized
Sch\"{o}dinger's cat state, but also provides the platform to
exploit quantum metrology and quantum information processing through
ultrafast optics.

This work is supported in part by the Ministry of Science and
Technologies, Taiwan, under the contract No. 101-2628-M-007-003-MY4,
 No.  103-2221-E-007-056, and No. 103-2218-E-007-010.

\end{document}